\documentclass{emulateapj}
\usepackage{amssymb}
\usepackage{amsmath}

\input{psfig}
\slugcomment{Accepted for publication in the Astrophysical Journal}
\shorttitle{Cold accretion and evolving IMF}

\begin{document}

%%%%%%%%%%%%%%%%%%%%%%%%%%%%%%%%%%%%%%%%%%%%%%%%%%%%%%%%%%%%%%%%%%%%%%%%%%

\title{Galaxy formation with  cold  gas   accretion  and   evolving stellar  initial   mass  function}
\author {Xi Kang$^{1,2}$, W. P. Lin$^{3}$, Ramin Skibba$^{2,4}$, D. N. Chen$^{5}$}
\affil{1 The Purple Mountain Observatory, 2 West Beijing Road, Nanjing 210008, China}
\affil{2 Max-Planck-Institute for Astronomy, K\"onigstuhl 17, D-69117
  Heidelberg, Germany}
\affil{3 Key Laboratory for Research  in Galaxies and Cosmology, Shanghai
  Astronomy Observatory, Nandan Road 80, Shanghai, 200030, China}
\affil{4 Steward Observatory, University of Arizona, 933 N.Cherry Avenue, Tucson, AZ 85721, USA} 
\affil{5 Beijing Planetarium, No 138 Xizhimenwai Street, Beijing 100044, China}

\email{kangxi@pmo.ac.cn}
%%%%%%%%%%%%%%%%%%%%%%%%%%%%%%%%%%%%%%%%%%%%%%%%%%%%%%%%%%%%%%%%%%%%%%%%%%

%\pubyear{2000}

%%%%%%%%%%%%%%%%%%%%%%%%%%%%%%%%%%%%%%%%%%%%%%%%%%%%%%%%%%%%%%%%%%%%%%%%%%

\begin{abstract} 

  The  evolution of  the galaxy  stellar mass  function  is especially
  useful to test the current model of galaxy formation.  Observational
  data have  revealed a few inconsistencies with  predictions from the
  $\Lambda {\rm CDM}$ model.   For example, most massive galaxies have
  already  been  observed  at  very  high  redshifts,  and  they  have
  experienced only mild evolution  since then.  In conflict with this,
  semi-analytical models  of galaxy formation  predict an insufficient
  number of  massive galaxies at  high redshift and a  rapid evolution
  between  redshift  1  and  0  .   In addition,  there  is  a  strong
  correlation  between  star  formation  rate  and  stellar  mass  for
  star-forming  galaxies, which  can  be roughly  reproduced with  the
  model, but  with a normalization that  is too low  at high redshift.
  Furthermore, the stellar mass  density obtained from the integral of
  the cosmic star formation history is higher than the measured one by
  a  factor of  2.   In this  paper,  we study  these  issues using  a
  semi-analytical  model  that  includes:  1) cold  gas  accretion  in
  massive halos at  high redshift; 2) tidal stripping  of stellar mass
  from  satellite galaxies; and  3) an  evolving stellar  initial mass
  function  (bottom-light) with  a higher  gas recycle  fraction.  Our
  results show  that the combined effects  from 1) and  2) can predict
  sufficiently massive galaxies at  high redshifts and reproduce their
  mild evolution at low redshift, While the combined effects of 1) and
  3)  can reproduce the  correlation between  star formation  rate and
  stellar  mass  for  star-forming   galaxies  across  wide  range  of
  redshifts.   A  bottom-light/top-heavy   stellar  IMF  could  partly
  resolve  the conflict between  the stellar  mass density  and cosmic
  star formation history.

\end{abstract}

%%%%%%%%%%%%%%%%%%%%%%%%%%%%%%%%%%%%%%%%%%%%%%%%%%%%%%%%%%%%%%%%%%%%%%%%%%


%%%%%%%%%%%%%%%%%%%%%%%%%%%%%%%%%%%%%%%%%%%%%%%%%%%%%%%%%%%%%%%%%%%%%%%%%%

\section{Introduction}
\label{sec:intro}

The current  paradigm of  structure formation based  on the  cold dark
matter (CDM)  model can  explain the pattern  of mass  distribution on
large  scales  (e.g.,  Springel  et  al.   2006).   In  this  picture,
structure ( consisting  of dark matter halos) forms  in a hierarchical
manner  such  that small  halos  tend  to  form relatively  early  and
subsequently merge to form large halos.  Galaxies are expected to form
at the  center of dark  matter halos via  star formation in  clouds of
cooling  gas  (White  \&  Rees   1978).   In  the  last  few  decades,
semi-analytical models  (SAMs) of galaxy formation,  which combine the
formation history  of dark matter  halos and physical  descriptions of
star  formation  and feedback  mechanisms,  have  been established  as
useful  tools  for investigating  the  impacts  of different  physical
processes on galaxy properties.  Overall, SAMs can be constructed such
that they agree with a wide rage of observations.  In particular, with
realistic models (including supernova  and AGN feedback), the SAMs can
match  the faint  and luminous  end of  luminosity functions  of local
galaxies well (see Baugh et al. 2007 for a review).

Recently,  observations of  galaxies  at high-$z$  have revealed  some
issues that are difficult to explain with the model. Most importantly,
the data suggest that the formation of galaxies is 'anti-hierarchical'
(sometimes referred to as 'downsizing'; see Cowie et al.  1996).  Most
massive galaxies  were already  in place at  very early  epochs (e.g.,
Pozzetti  et al.  2003),  while low-mass  galaxies were  still forming
with modest star formation  rates.  Measurements of the galaxy stellar
mass function also show that massive galaxies evolve relatively little
from $z$=1 to  $z$=0, whereas the SAMs predict  an insufficient number
of  massive galaxies  at high-$z$  (e.g., Kitzbichler  \&  White 2007;
Fontanot et al. 2009) and  the predicted evolution of massive galaxies
is too rapid compared to the observations (e.g., De Lucia et al. 2007)
unless the  disruptions of satellite  galaxies are invoked  (Monaco et
al.  2006; Conroy et al. 2007; Somerville et al. 2008).

The absence of massive galaxies at  early times in the SAMs is related
to the modeling of gas cooling (e.g., Yoshida et al. 2002; Cattaneo et
al.  2006).  In  most SAMs it is assumed that  hot gas is isothermally
distributed and cools down via  radiative cooling, as described by the
classical model  of White \&  Frenk (1991).  However, it  was recently
found  that gas  cooling  is  dominated by  cold  accretion, which  is
efficient in  halos with mass below $10^{12}M_{\odot}$,  where the gas
pressure  is unable to  support a  stable shock  (e.g., Kere{\v  s} et
a. 2005).  In addition, cold accretions are also efficient in high-$z$
massive halos, as cold gas can flow along filaments (Birnboim \& Dekel
2003; Dekel et al. 2009).

Star-forming galaxies have a strong correlation between star formation
rate  and stellar  mass (SFR-$M_{\ast}$  relation), and  this relation
evolves  strongly  with  redshift  (e.g., Brinchmann  \&  Ellis  2000;
Brinchmann et al. 2004; Feulner et  al. 2005; Daddi et al. 2007; Elbaz
et  al.  2007;  Noeske et  al.   2007; Zheng  et al.   2007; Drory  \&
Alvarez 2008;  Dunne et al.   2008). SAMs can approximately  match the
observed SFR-$M_{\ast}$ relation (Kitzbichler \& White 2007), but with
a normalization that is too low at $z>0$ (Daddi et al.  2007; Elbaz et
al.   2007).   Khochfar  \&  Silk  (2009)  recently  showed  that  the
inclusion of  cold accretion can boost the  SFR-$M_{\ast}$ relation to
agree with the data at $z$=2.

The  observed  SFR-$M_{\ast}$ relation  is  dependent  on the  assumed
stellar  initial  mass  function  (IMF). Dav\'{e}  (2008)  proposed  a
stellar IMF, which is bottom-light  compared to the Kroupa IMF (Kroupa
2001) at  high-$z$, and they  find that the observed  normalization of
SFR-$M_{\ast}$  relation can  be reduced  with a  carefully determined
stellar IMF.  An evolving IMF is not a naive speculation, although its
exact form is far from  clear.  There is theoretical and observational
evidence  for  the  stellar  IMF   at  high-$z$  to  be  top-heavy  or
bottom-light (e.g., Larson 2005; van Dokkum 2008).  Constraints on the
stellar  IMF  can  also  be  obtained  from  comparisons  between  the
evolution of cosmic star formation rate and stellar mass density.  For
example, Wilkins  et al.  (2008)  found that the stellar  mass density
derived  from the  cosmic star  formation history  is higher  than the
measured one by  a factor of $\sim 2-3$ with the  use of standard IMFs
(Salpeter  1954;  Kroupa 2001;  Chabrier  2003).   However, some  have
argued that  this discrepancy may be  due to a dust  effect (Driver et
al.   2007) and incompleteness  in the  measured stellar  mass density
(Reddy \& Steidel 2009).  A top-heavy  IMF is also favored in order to
provide enough photons  to ionize the universe at  $z>7$ (e.g., Bunker
et al. 2009).

In this paper,  we use the model  of Kang et al.  (2005)  to study the
evolution  of  the stellar  mass  function,  luminosity function,  and
SFR-$M_{\ast}$  relation. We  modify  the model  to  include cold  gas
accretion, and we use the  evolving IMF proposed by van Dokkum (2008).
In   Section~\ref{sec:SAM},   we   introduce   the   model   and   its
modifications.   In Section~\ref{sec:SMF},  we  compare the  predicted
stellar mass  and luminosity  functions to the  data.  We  discuss the
SFR-$M_{\ast}$  relation in  Section~\ref{sec:sfr-ms} and  present the
evolution  of star  formation history  in  Section~\ref{sec:sfrh}.  We
briefly conclude the paper in Section~\ref{sec:concl}.

\section{Semi-analytical model}

\subsection{N-body Simulation}

We construct  halo merger trees from a  cosmological N-body simulation
with cosmological  parameters $\Omega_{m}=0.24, \Omega_{\Lambda}=0.76$, and 
$\sigma_{8}=0.76$,  which are  similar to  the WMAP3  cosmology results
(Spergel  et  al.  2007).   The  simulation  was  made using  Gadget-2
(Springel  2005)  and follows  $512^{3}$  particles  of mass  $5\times
10^{8}M_{\odot}/h$  in a  cube box  of  $100$ Mpc/h on  each side.   We
locate  dark  matter  halos  in  the  simulation  using  the  standard
friends-of-friends  (FOF) method,  and  identify subhalos  in the  FOF
halos.   We then  construct the  subhalo  merger trees  and apply  the
physical model  to describe star formation within  these subhalos (see
Kang et al. 2005 for details).

\subsection{Model for galaxy formation}
\label{sec:SAM}

The model  used in this  paper is based  on Kang et al.   (2005; 2006;
Kang  2009).  Kang  et al.   (2005) show  in detail  how to  graft the
galaxy  formation model  onto the  subhalo-resolved merger  trees.  In
Kang et  al.  (2006), an AGN  feedback model was  included to suppress
gas cooling in  massive halos.  Such a feedback  mechanism is found to
reproduce  the  luminosity of  central  cluster  galaxies well  (e.g.,
Croton et  al.  2006;  Bower et al.   2006; Somerville et  al.  2008).
The  model   is  further  improved   by  Kang  (2009)  to   include  a
photo-ionization  model  to  suppress  the  formation  of  very  faint
galaxies.  Photo-ionization  is essential to  reproduce the luminosity
function of Milky-Way satellites ( Macci{\`o} et al.  2009).  We refer
the reader to these papers for more details.

 In the following, we will introduce the modifications to the previous
 version  of the  model, including  a simple  description of  cold gas
 accretion,  an evolving  IMF,  and  a more  realistic  dust model  to
 describe the extinction in galactic disks and starburst galaxies.  We
 also  follow Somerville  et al.   (2008)  to model  the stellar  mass
 stripping of satellite galaxies.

 \begin{itemize}

 \item  We  include  a  simple  description for  cold  gas  accretion,
   following  results from  simulations.  In  previous models,  gas is
   assumed to be shock-heated to  the halo virial temperature and then
   cool down via radiation.  The cooling rate is described by Equation
   3-5 of  Kang et al.   (2005) and Equation  3 of Kang et  al (2006).
   Here, we  refer to this cooling description  as hot-mode accretion.
   Kere{\v s} et  al.  (2005; 2008) have shown  that cold-mode accretion is
   expected in  halos of mass below $3-10  \times 10^{11}M_{\odot}$ (a
   threshold that depends on metallicity),  as the gas is never heated
   to the halo virial temperature.   Dekel \& Birnboim (2006; Dekel et
   al.    2009)    have   shown   that   even    for   massive   halos
   $M>10^{12}M_{\odot}$  at high$-z$, cold  streams still  dominate, as
   the gas is infalling along  filaments that are thin compared to the
   halo virial radius and hence avoids being shock heated.  Both phases
   of gas accretion can occur simultaneously, and their fraction varies
   with halo mass and redshift (Ocvirk et al. 2008).

   Dekel et al.  (2009) have shown that the gas accretion is dominated
   by  cold flows within  halos below  a critical  mass, which  can be
   described as  $M_{c}=M_{0}max[1, 10^{1.1(z-z_{c})}]$.  Depending on
   the  metallicity of  the gas  , $M_{0}$  is about  $\sim3-20 \times
   10^{11}M_{\odot}$, and  $z_{c}$ is  between $\sim 1.3-2$.   Here we
   use $M_{0}=10^{12}M_{\odot}$ and $z_{c}=2.0$.  All available gas in
   halos with  mass below $M_{c}$ is  assumed to be  accreted into the
   central disk  by the halo dynamical time-scale.   For massive halos
   ($M > M_{c}$), we continue to use hot-mode accretion.

   In  general,  the  cold  accretion  model  allows  more  rapid  gas
   accumulation  in  massive  halos  at  high-$z$.  We  will  show  in
   Section~\ref{sec:SMF}  that compared to  the hot-mode  accretion it
   produces  more massive  galaxies  at $z>2$  and  predicts a  larger
   number of galaxies with high star formation rates.

 \item All previous SAMs use constant stellar IMFs.  In this paper, we
   consider  two  possible  stellar  IMFs.   The first  is  the still 
   frequently used  Chabrier IMF  (Chabrier 2003, hereafter  C03). The
   other is the  evolving IMF proposed by van  Dokkum (2008, hereafter
   VD08).  Although there are  no direct observational measurements of
   the  stellar IMF  at high-$z$,  there is  emerging evidence  for an
   evolving  IMF from  both theoretical  and  observational arguments.
   Theoretically, Larson  (2005; also  Jappsen et al.   2005) suggests
   that the  characteristic mass of  star formation is related  to the
   minimum  temperature  of  molecular  clouds, which  increases  with
   redshift   due  to   heating  from   cosmic   microwave  radiation.
   Observationally,  some have  also favored  an evolving  stellar IMF
   (e.g.,  Lucatello et  al.  2005;  Stanway et  al. 2005;  Hopkins \&
   Beacom 2006; Tumlinson 2007; Wilkins et al.  2008; Elmegreen et al.
   2008; Bunker et al. 2009).  For example, Wilkins et al. (2008, also
   Dav\'{e} 2008) have shown that a standard stellar IMF (Salpeter IMF
   or Chabrier  IMF) produces a  higher stellar mass density  than the
   observed one.

   Recently,  van Dokkum  (2008) studied  the color  and mass-to-light
   ratio of  elliptical galaxies between  $0<z<1$, and found  that the
   data favor  an IMF with fewer  low-mass stars at $z  \sim 4$.  They
   used a  modified Chabrier  IMF to  fit the IMF  data of the Milky Way
   disk, globular  cluster, submm-galaxies (see their  Equation 18 and
   20).   For the VD08  IMF, its characteristic  mass $m_{c}$
   evolves  with redshift, such  that at  $z$=0 it  is $0.08M_{\odot}$
   (C03 IMF), but  increases to $1M_{\odot}$ at $z$=5.  Thus, the VD08
   IMF includes more high-mass stars at high redshifts.

   Observationally, the star formation rates and stellar mass are both
   dependent on the assumed stellar IMF.  Therefore, the data should
   be corrected if a different stellar IMF is used.  For this purpose,
   we use the stellar population synthesis code of PEGASE.2 (Fioc \&
   Rocca-Volmerange 1997).  We refer the reader to van Dokkum (2008)
   for details about how to correct the SFR and stellar mass. The SFR
   correction is relatively simple because it is only dependent on the
   number of massive stars.  The stellar mass correction is not
   straightforward, as it is also dependent on the assumed star
   formation history.

  An important  effect from  varying IMF is  that the  model parameter
  $R$,  the gas  recycle  fraction from  evolving  stars, should  also
  change.   Following  most SAMs,  we  calculate  $R$  from a  stellar
  population   synthesis  model,  using   the  PEGASE.2.   Wilkins  et
  al. (2008)  have shown that different models  produce similar values
  of $R$. An  IMF consisting of more high-mass  stars will return more
  gas  into  the interstellar  medium,  and  produce  a lower  stellar
  remnant  mass.  We found  that higher  $R$ will  result in  a higher
  normalization of  the SFR-stellar mass correlation.  Also higher $R$
  is  helpful to solve  the discrepancy  between the  measured stellar
  mass  density  and that  inferred  from  the  cosmic star  formation
  history.

 \item The tidal  stripping of galaxy stellar mass  is observed in the
   Milky Way (Ibata et al.  2001) as well as in galaxy clusters (e.g.,
   Mihos et al.  2005; Zibetti  et al.  2005).  N-body simulation is 
   a useful  tool for studying this process,  although currently there
   is a lack of  detailed studies  about how  the  stripping efficiency
   depends  on the  properties of  the satellite  galaxy and  its host
   halo.  Here  we follow Somerville  et al.  (2008) by  assuming that
   before  merging with central  galaxies, a  fraction $f_{s}$  of the
   stellar mass  of satellite galaxies is stripped  and scattered into
   the intracluster  region. We take  their value of 0.4  for $f_{s}$,
   and we  find that the intracluster  light consists of  about 8\% of
   the total cluster light in  our model, which is consistent with the
   results of Zibetti  et al. (2005). Due to  the difficult-to-measure
   low  surface brightness objects, the measured fraction  of intracluster
   light spans a wide range between  5-50\% (see review by Lin \& Mohr
   2004; Feldmeier et al. 2004).

 \item A dust extinction model similar  to that of Guo \& White (2009)
   is used  in this  paper.  Compared to  the previous version  of our
   model, it accounts for the  dependence of optical depth on the cold
   gas fraction and metallicity.  We  apply this dust extinction model to
   galaxy disks.  For galaxies experiencing recent starbursts, we use
   the  dust model  of  Calzetti  et al.   (2000)  instead, and  adopt
   $E(B-V)$  as 0.3,  which  is typical  for  dusty starburst  galaxies
   (Calzetti et  al.  2000; Poggianti  \& Wu 2000).  In  the resulting
   model, we  find that  at $z=0$ most  starburst galaxies  reside in
   halos with mass around $10^{12}M_{\odot}$.
\end{itemize}

\section{Stellar mass and Luminosity functions}
\label{sec:SMF}

\begin{figure*}
\centerline{\psfig{figure=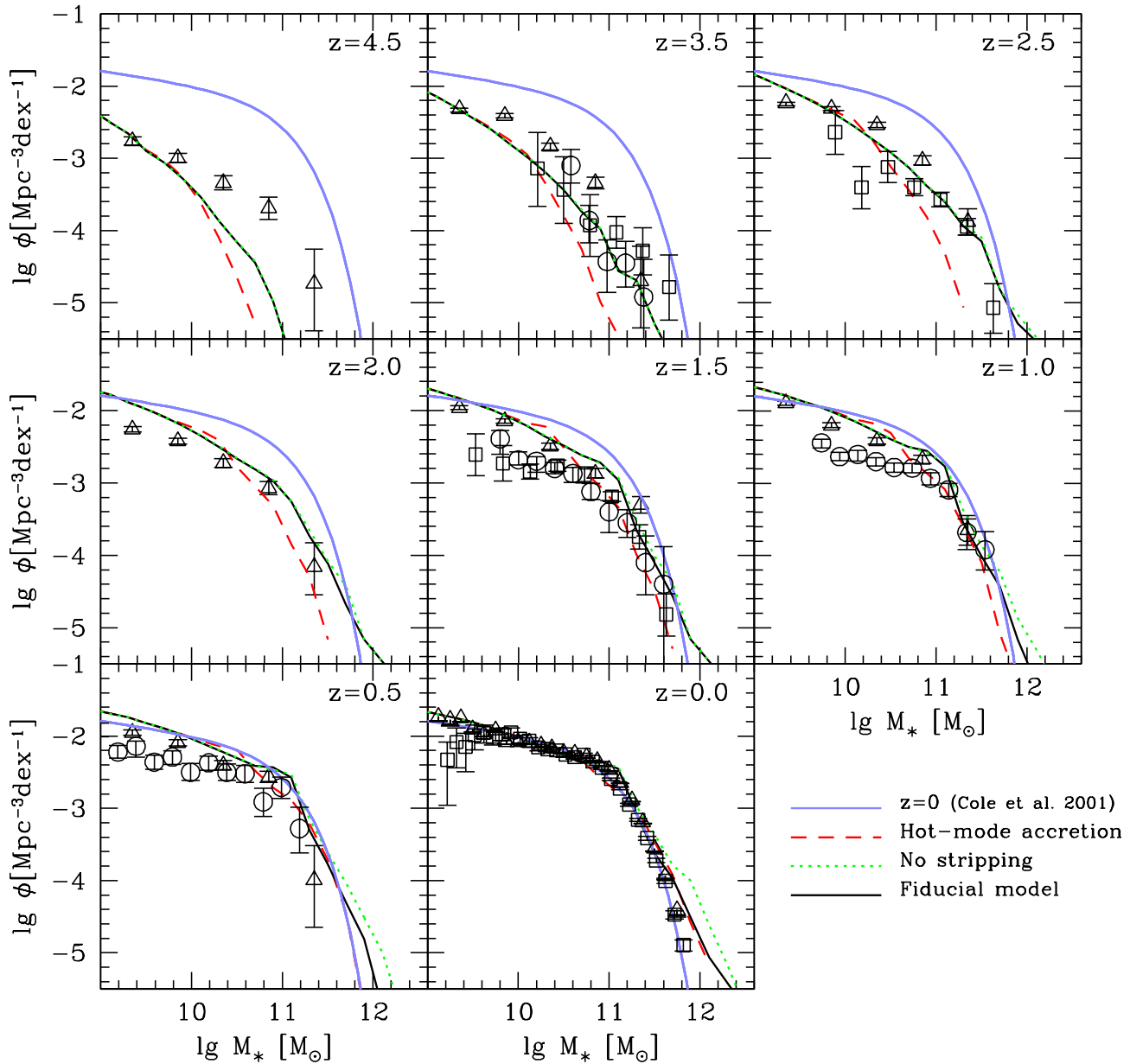,width=0.95\hsize}}
\caption{The  evolution of  galaxy stellar  mass functions.   The data
  points at z=0  are taken from Cole et al.   (2001, squares), Bell et
  al. (2003, triangles).  The high-z data are taken  from Drory et al.
  (2005, triangles),  Fontana et al.  (2006,  circles), and Marchesini
  et al. (2009, squares). The best fit  to the local SMF by Cole et al. is
  duplicated as gray line  in each panel. The red-dashed, green-dotted and
  black-solid  lines   are  our  model   predictions  under  different
  assumptions,  see  the main  text.   Note  that  here all  data  are
  transferred into the Chabrier 2003 IMF.}
\label{fig:smf}
\end{figure*}

In this  section, we compare the  predicted stellar mass  function (SMF )
and  rest-frame   $B$-  and  $K$-band  luminosity   functions  to  the
observational measurements.  The SMF  has been extensively measured in
 recent  decades using multi-wavelength photometry  data from large
area surveys.   The uncertainties in the measured  stellar mass result
from  the  adopted  stellar  IMF, dust  correction,  metallicity,  and
assumed star formation history, among which the stellar IMF contributes
the  dominant systematic  error between  different data  sets.   It is
simple to  correct the stellar  mass between the often  used Salpeter,
Kroupa  (2001) and  C03 IMFs,  which are  assumed to  be  independent of
galaxy mass  and redshift (Bell  et al. 2003).  As  the characteristic
mass of the VD08 IMF  evolves with time, the correction does depend 
on galaxy mass and redshift (Marchesini et al.  2009), and determining
these  dependences  is  beyond  the  scope  of  this  paper.  For  our
comparison of galaxy SMFs, we continue to use the C03 IMF and focus on
the  effects  of cold  accretion  and  tidal  stripping on  the  model
predictions.

In  Fig.~\ref{fig:smf}, we  plot the  data from  Drory et  al.  (2005,
triangles),  Fontana  et  al.    (2006,  circles)  and  Marchesini  et
al. (2009, squares)  at $z>0$, and the data at  $z\sim0$ are from Bell
et al.   (2003, triangles) and Cole  et al. (2001,  squares). They are
all converted to  the C03 IMF.  The  best fit to the data  at $z=0$ by
Cole et al.  (2001) is duplicated in each plot by the gray solid line,
and the black and color lines are our model predictions.

\begin{figure*}
\centerline{\psfig{figure=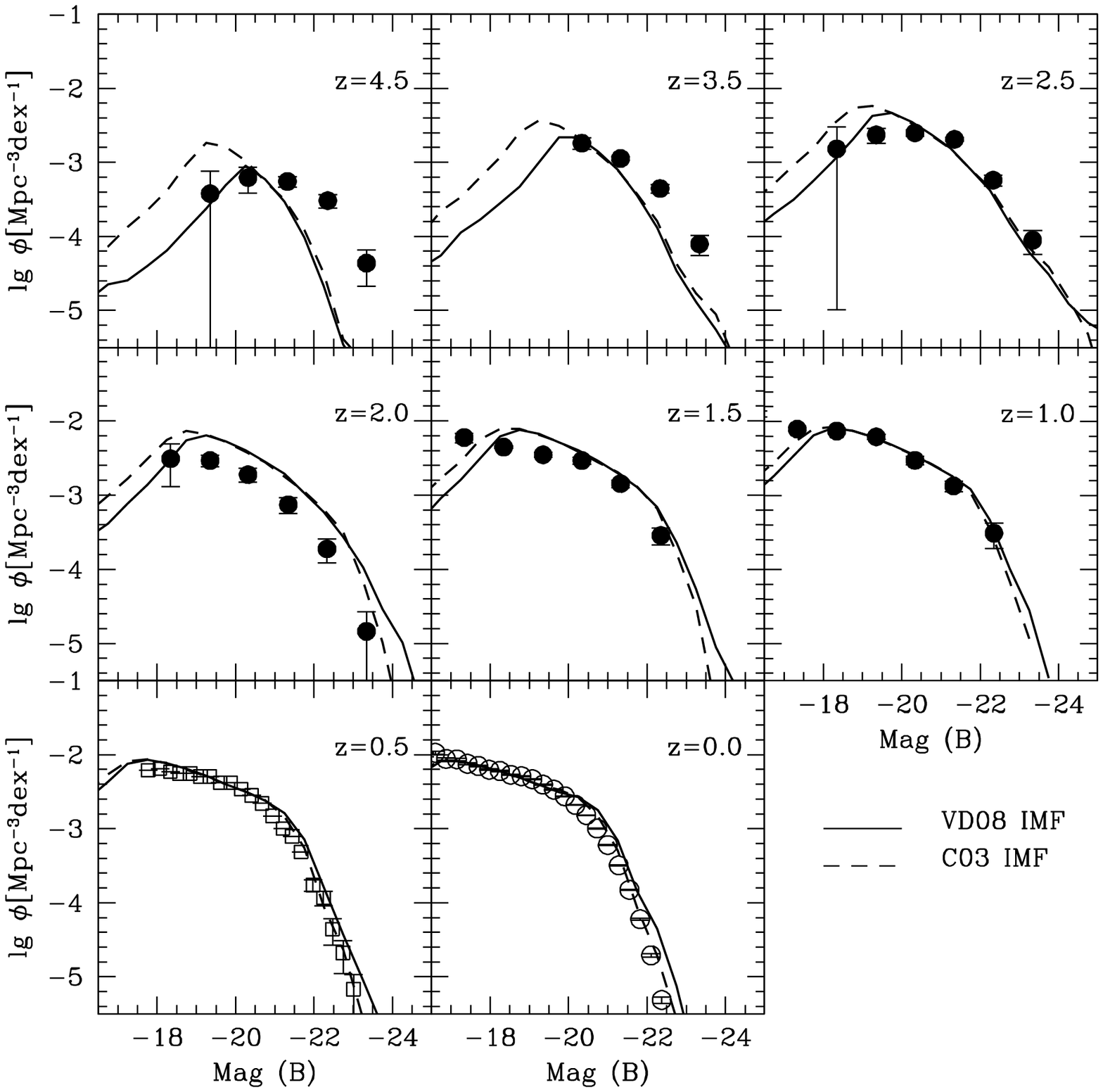,width=0.45\hsize}}
\caption{The evolution of rest-frame  B band (AB magnitude) luminosity
  functions.  Data points  are taken from Norberg et  al. (2002, z=0),
  Wolf et  al. (2003,  squares at z=0.5),  and Gabasch et  al.  (2004,
  solid  circles at  $z>0.5$). Solid  and dashed  lines are  our model
  predictions for two  stellar IMFs. Here we also  use dust extinction
  to young star-burst galaxies, see the text.}
\label{fig:lfb}
\end{figure*}

\begin{figure*}
\centerline{\psfig{figure=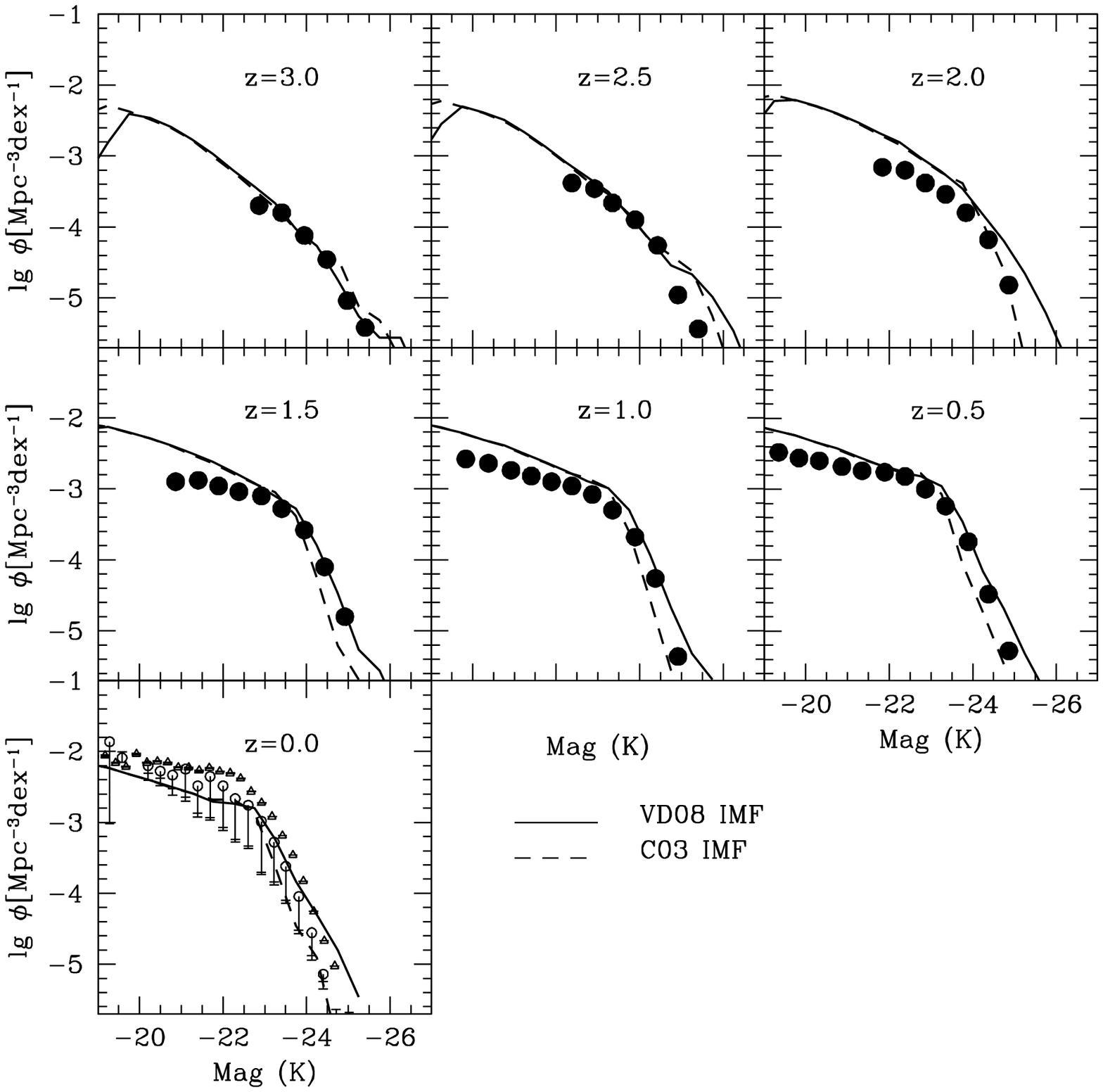,width=0.45\hsize}}
\caption{As  in  Fig.2,  but  for  the rest-frame  K  band  luminosity
  functions. Data points are taken from Cole et al. (2001,triangles at
  z=0),  Huang  et  al.  (2003,  circles at  z=0),  and  Cirasuolo  et
  al.  (2007, solid  circles). Note  that the  rest-K  band luminosity
  functions are shown up to z=3, lower than the B-band LFs in Fig.2}
\label{fig:lfk}
\end{figure*}

Firstly, it is clear that while the data are consistent at the massive
end, the  measurements differ  by a factor  of $\sim2-3$  for galaxies
with  mass  below $10^{10.5}M_{\odot}$.   This  is  mainly  due to  an
incomplete analysis of the errors. Most work consider only the Poisson
errors or errors from photometric redshift uncertainties. In addition,
field-to-filed  variations  are also  a  significant  source of  error
(Marchesini  et al.   2009).  The  green dotted  lines show  the model
prediction without tidal stripping.  As one can see, between $z$=1 and
$z$=0, this  model predicts  a stronger evolution  at the  massive end
than indicated by the data (Also see Somerville et al. 2008).  The red
dashed lines show  the prediction for hot-mode accretion,  and we find
that it produces a good match  to the local stellar mass function, but
predicts too few massive galaxies at $z>2$.  Our fiducial model, which
includes both tidal stripping and  cold gas accretion, produces a fair
match to  the stellar  mass function from  $z=4$ to $z=0$.   Still, we
find that  all of the SAMs  predict an overabundance  of galaxies with
mass  around  $10^{11}M_{\odot}$  at  $z \sim  1$.   Nonetheless,  the
inclusion of  cold gas  accretion is helpful  to produce  more massive
galaxies at high redshift, while the tidal stripping can reproduce the
mild evolution  of the stellar mass  function at the  massive end from
$z=1$ to $z=0$.

Although the  galaxy stellar mass  function is important  to constrain
the SAMs,  determining the galaxy stellar  mass involves uncertainties
from the modeling.  Marchesini et al.  (2009) have analyzed the varies
effects, such as dust,  metallicity, and stellar population synthesis,
on the measured  stellar mass.  For example, they  have found that the
stellar mass  derived from the Maraston  (2005) model is  lower by 0.2
dex  compared to  that obtained  using the  Bruzual \&  Charlot (2003)
model, and it is weakly dependent  on mass and redshift. In this case,
we argue that  the stellar mass correction has  no significant effects
on our results,  as we have normalized our  model parameters using the
local stellar mass function. It is the evolution of the galaxy stellar
mass function that puts strong constraints on our model parameters.

We show the  evolution of the rest-frame $B$-  and $K$-band luminosity
functions   (LFs)   in   Fig.~\ref{fig:lfb}  and   Fig.~\ref{fig:lfk},
respectively.  The  $B$-band LF is  more sensitive to the  recent star
formation  activity of galaxies,  rather than  the total  stellar mass
indicated by the  $K$-band LF.  Measurements of LFs  at both bands can
set   constraints  on   the  star   formation  history   of  galaxies.
Observational data  have improved rapidly over  recent years. However,
due to  the magnitude  limit and  survey volume, the  LFs at  both the
faint and luminous ends are  not well constrained, but in general, all
of these measured LFs are consistent with each other.

\begin{figure*}
\centerline{\psfig{figure=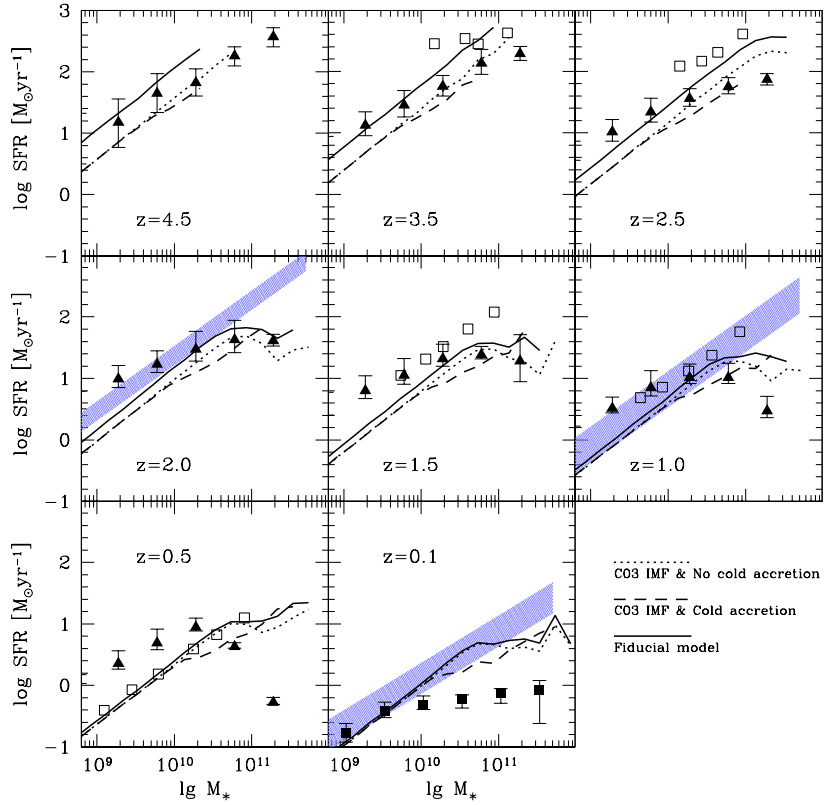,width=0.6\hsize}}

\caption{The star formation rate  - stellar mass relation from $z=4.5$
  to  $z=0$.  Data  points  are  taken from  Drory  \& Alvarez  (2008,
  triangles),  Dunne et  al. (2009,  empty squares),  and Chen  et al.
  (2009,  solid squares).  Shade  area are  best fits  (with 1$\sigma$
  variance) by  Daddi et al.  (2007)  and Elbaz et al.   (2007) to the
  data at  $z=2,1,0$. Our  model predictions are  shown as lines
  with model parameters indicated in the plot.}
\label{fig:sfr-ms}
\end{figure*}

The  $B$-band data  are taken  from Norberg  et al.   (2002),  Wolf et
al. (2003),  Gabasch et  al (2004). The  $K$-band data are  taken from
Cole et al. (2001), Huang et  al. (2003), and Cirasuolo et al. (2007).
Our model predictions are shown as  the solid and dashed lines for the
VD08  and C03  IMFs,  respectively.  Overall  we  find good  agreement
between  the model  predictions and  the data  in both  $B$-  and $K$-
bands.   The agreement at  high redshifts  is encouraging  as previous
models  (e.g., Kitzbichler  \& White  2007; Bower  et al.   2006) have
under-predicted the  abundance of bright galaxies in  $K$-band at high
redshifts.  As   stated  previously,  this   is  mainly  due   to  the
implementation of cold accretion in  the model. Note that the $B$-band
LFs drop off at the faint end at high redshift, and this is due to the
resolution of  our simulation,  in which only  halos with  mass larger
than $5\times 10^{9}M_{\odot}/h$ are included.

We also  find that the model  predictions only slightly  depend on the
choice of IMF.  This is unexpected, because the VD08 IMF contains more
high-mass  stars.   However,  as   shown  by  van  Dokkum  (2008)  and
Marchesini et  al.  (2009), although  there are more massive  stars in
the VD08  IMF, the  number of characteristic  stars that  dominate the
optical light, also decreases  with increasing redshift.  As a result,
the net effects on the luminosity are modest.

\section{The star formation rate - stellar mass relation}
\label{sec:sfr-ms}

We investigate the SFR-$M_{\ast}$  relation in this section.  A direct
comparison between the model  and the observational data is difficult,
because of  the various selection  effects in the data.   Firstly, the
tight  correlation between  SFR and  $M_{\ast}$ is  observed  only for
star-forming galaxies, and  there lacks of a simple  and robust way to
select  star-forming  galaxies  in  the  model to  match  the  various
observational samples.   Secondly, due to survey  limits, many low-SFR
(or passive) galaxies are not included in observational samples, which
may give rise to a biased SFR-$M_{\ast}$ relation.  However, given the
tightness of  the SFR-$M_{\ast}$ relation  and its small  scatter, the
missing  galaxies  can  not  contribute significantly  above  the  SFR
detection limit (Dav\'{e} et al. 2008).

We  select  all the  model  galaxies and  plot  their  average SFR  in
Fig.~\ref{fig:sfr-ms}.  The  data plotted  are Drory \&  Alvarez 2008;
triangles), Dunne et al. (2009;  open squares), and Chen et al. (2008;
filled squares).  The  shaded area show the best  fits (with $1\sigma$
variance) to  the data at $z=2, 1,  0$ quoted by Daddi  et al.  (2007)
and Elbaz et al  (2007).  It is important to note that  the Chen et al
(2008) and  Drory \& Alvarez  (2008) data are  the average SFR  of all
galaxies in their samples, while the  other data are only for the main
sequences  of active  star-forming  galaxies or  sBzK galaxies.   This
gives rise  to the different  behavior of the  SFR-$M_{\ast}$ relation
for massive  galaxies, where the  average SFR of all  galaxies departs
from the sBzK result because  more massive and passive galaxies become
dominant at  lower redshift.   For less massive  galaxies, we  can see
that the data agree with each other within a factor of $\sim2$.

Our model predictions are shown  by the solid, dashed and dotted lines
in Fig.~\ref{fig:sfr-ms}.  Our fiducial model, with cold accretion and
an  evolving  IMF, can  reproduce  the  evolution  of the  correlation
between SFR  and $M_{\ast}$ fairly well.  Although  our predictions do
not  agree with all  of the  measurements, they  are within  the range
spanned  by different  data  sets.  For  massive  galaxies, our  model
predictions  are in  better agreement  with Drory  \&  Alvarez (2008),
because both are  the average SFR of all  galaxies.  At lower redshift
($z < 1$), our SFR-$M_{\ast}$ relation has a higher tail than the data
at the massive end.  In general, our predicted SFR-$M_{\ast}$ relation
is a  significant improvement over previous results  from SAMs (Croton
et  al.   2006;  Kitzbichler  \&  White  2007),  which  have  produced
SFR-$M_{\ast}$ relations that are too low compared to the observations
at  $z=2$ and  $z=1$.   Recently  Fontanot et  al.   (2009) have  also
examined the SFR-$M_{\ast}$ relation,  but their predictions are still
lower than  the observations.  In  the following, we  will investigate
possible contributions to the  boost of the SFR-$M_{\ast}$ relation in
our model vis-\'{a}-vis other models.

\begin{figure}
\centerline{\psfig{figure=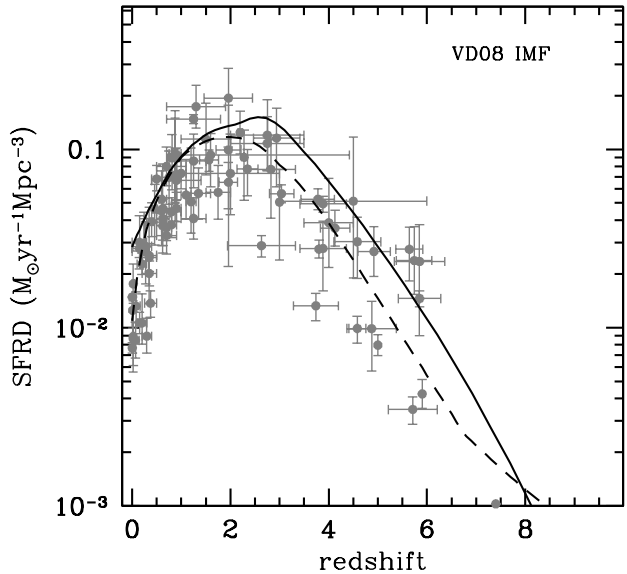,width=0.95\hsize}}
\caption{The cosmic  star formation rate history. The  gray points are
  taken  from the  compilation  of observational  data  by Hopkins  \&
  Beacom (2006),  and the  dashed line  is the best  fit to  the data,
  which are  all corrected to the  VD08 IMF.  Our  model prediction is
  plotted as the solid line.}
\label{fig:sfrh}
\end{figure}

For the model  with the traditional hot-mode cooling  and C03 IMF, the
prediction is in  agreement with the data at $z<1$,  but is lower than
the  measurements  at  $z=2$.   This  is similar  to  the  results  of
Kitzbichler \& White (2007) and  Fontanot et al.  (2009). In contrast,
when cold accretion is included, the model predicts more galaxies with
high star formation  rates, so that the SFR-$M_{\ast}$  relation is in
better agreement  for $M_{\ast} >  10^{10}M_{\odot}$, but it  is still
lower that the data at $z>2$.

The  normalization of  the predicted  SFR-$M_{\ast}$ relation  is also
boosted in our fiducial model by the VD08 IMF. This boost is larger at
higher redshift, but is independent of galaxy mass.  In the model with
the  VD08 IMF,  at high  redshift there  are more  massive  stars that
return more stellar winds to the interstellar medium and thus decrease
the stellar remnant  mass. For example, at $z=4$,  the stellar remnant
at age  of 0.5 Gyr is  28\% of the  initial formed mass with  the VD08
IMF, but it is  63\% for the C03 IMF. Thus for  a given star formation
rate  history, the  stellar  mass is  lower  for the  VD08 IMF,  which
significantly boosts the  normalization of SFR-$M_{\ast}$ relation. We
find that  the SFR-$M_{\ast}$  relation increases by  $\approx60\%$ at
$z=2$.

\begin{figure}
\centerline{\psfig{figure=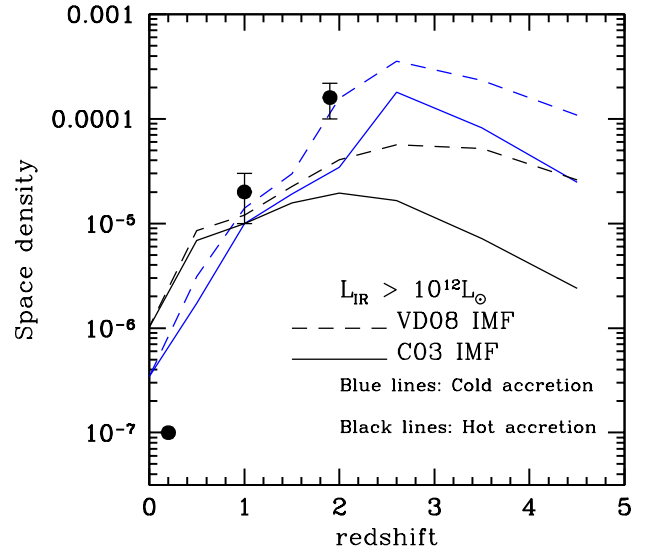,width=0.95\hsize}}
\caption{The    space    density   of    ULIRGs    with   $L_{IR}    >
  10^{12}M_{\odot}$. The implied star formation rate for these objects
  is  $SFR >  120M_{\odot}/yr$ for  the  Chabrier 2003  IMF, but  only
  $84M_{\odot}/yr$  for the VD08  IMF at  z=2,  which is
  bottom-light than the C03 IMF.  The  data point at z=2 implies that both
  cold  accretion and bottom-light  (or top-heavy)  IMF are  needed to
  produce enough ULIRGs.}
\label{fig:ULIRGs}
\end{figure}

Finally, we note that these results depend strongly on the gas recycle
fraction  $R$. As  noted previously,  $R$ is  calculated  from stellar
population synthesis,  which involves uncertainties  from the modeling
of stellar winds and stellar evolution tracks. More strong constraints
could be  put on the IMF  with more precise modeling  of stellar winds
and supernova feedback.

\section{Cosmic Star formation history}
\label{sec:sfrh}

Many recent  observations have been dedicated to  measuring the cosmic
star formation  history (e.g., Madau  et al.  1996; Giavalisco  et al.
2004;  Bouwens  et  al.  2006).   It  has  been  found that  the  star
formation rate density (SFRD) at redshift $1-2$ is higher than that of
the local universe by an  order of magnitude.  The precise location of
the peak of the  SFRD is not yet clear, but it  may be beyond $z\sim2$
(Steidel et al.  1999).  With  more data from submm galaxies and Lyman
break  galaxies,  the SFRD  can  be measured  out  to  the very  early
universe ($z \sim  7$).  For galaxy formation models,  it is important
to understand  which mechanism  drives the rapid  increase of  SFRD at
$z>2-3$, and the steep decrease at $z<1$.

As  our model  parameters are  normalized  by the  local stellar  mass
function, we  can make  predictions for the  history of the  SFRD.  In
Fig.~\ref{fig:sfrh}, we  compare the prediction of  our fiducial model
to observations at  $0<z<6$.  The data are taken  from the compilation
of observational data  by Hopkins \& Beacom (2006,  and see references
therein),  and we  add the  measurements  of Reddy  \& Steidel  (2008,
pentagon points) and Bouwens et al.  (2007, circles).  All of the data
and their best fit from  Hopkins \& Beacom (dashed line) are converted
to  the VD08  IMF.   We find  that  our model  reproduces the  overall
evolution of the SFRD quite well,  over a wide range of redshifts.  At
$z>3$ the data  are not yet well constrained,  although we expect that
new data in  the future will contribute additional  constraints on the
models.

Although  our  model predicts  the  total  SFRD  approximately, it  is
interesting to determine whether  it also predicts a sufficient number
of galaxies with very high  star formation rates.  The observations of
ultra-luminous  infrared  galaxies  (ULIRGs)  indicate that  they  are
possibly galaxies experiencing strong starbursts with dust obscuration
(Rieke  \&  Lebofsky  1979).  Daddi  et al.   (2007)  found  that  the
predicted  number   of  intensely  star-forming  galaxies   in  a  SAM
(Kitzbichler \& White  2007) is lower than the data at  $z$=1 and 2 by
an order of  magnitude.  Recently Khochfar \& Silk  (2009) showed that
the inclusion of cold accretion in their model produces enough ULIRGs.

In  Fig.~\ref{fig:ULIRGs},  we  show  our model  predictions,  and  we
compare them to the data from Sander et al. (2003, $z$=0) and Daddi et
al. (2007, $z>0$). The blue and black lines show the model predictions
with cold-mode and hot-mode accretion, respectively. We again consider
two possible IMFs (C03 IMF: solid lines; VD08 IMF: dashed lines).  The
ULIRGs with $L>10^{12}L_{\odot}$ have star formation rates larger than
$120M_{\odot}/yr$ for the  C03 IMF (for the VD08  IMF, the implied SFR
at $ z=2$ is $84M_{\odot}/yr$).

As can  be seen from the  figure, both cold accretion  and an evolving
IMF are  required for the  model to be  reconciled with the  data.  At
$z$=1, the  model cannot be constrained  well as cold  accretion is no
longer very important and the VD08  IMF is similar to the C03 IMF.  We
note that  data at $z>3$  contribute strong constraints on  the model,
and  additional constraints  will be  obtained from  observations from
future high-$z$ submm galaxy surveys.

\begin{figure}
\centerline{\psfig{figure=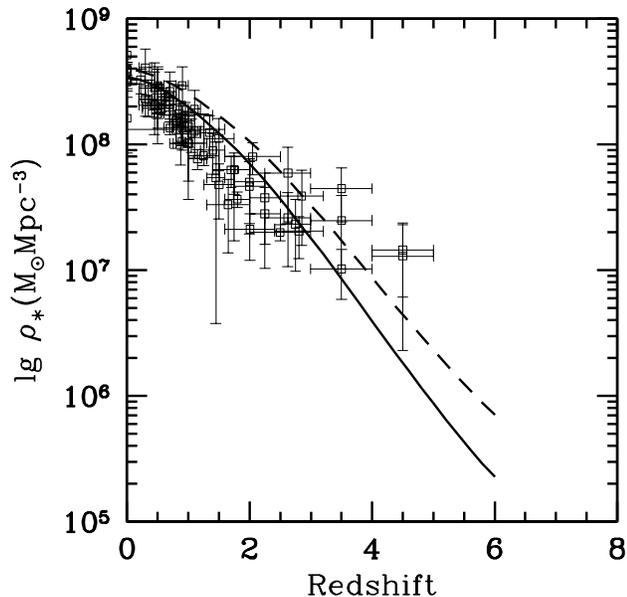,width=0.95\hsize}}
\caption{The evolution of stellar  mass density. Data points are taken
  from  the  compilation by  Wilkins  et  al.  (2008). The  lines  are
  integrated stellar mass density  from the cosmic star formation rate
  history (dashed line  in Fig.\ref{fig:sfrh}).  Different gas recycle
  fraction,$R$, are used  for the two lines. For  the dashed line, $R$
  is taken  from the C03 IMF, while  the solid line uses  $R$ from the
  VD08  IMF. We  can find  that better  agreement is  obtained  if the
  stellar remnant mass is lower, which is the case with a bottom-light
  (or top-heavy) IMF.}
\label{fig:sfrd}
\end{figure}

Finally,  we compare the  observed stellar  mass density  history with
that expected  from the integral  of cosmic star formation  history in
Fig.~\ref{fig:sfrd}. The integrated  stellar mass is simply calculated
as $\int SFR(t)(1-R)dt$, where $R$  is the fraction of gas returned to
the  inter-stellar medium by  stellar winds  and supernovae,  which is
dependent on the IMF and stellar  ages.  The data points in the figure
show  the measured  stellar mass  density compiled  by Wilkins  et al.
(2008),  and the  lines are  inferred from  the cosmic  star formation
history (the same as the dashed line in Fig.~\ref{fig:sfrh}), but with
different parameter $R$.

Some have found that the inferred stellar mass density from the cosmic
star formation history is higher than the measured one at $1<z<3$ with
either a Salpeter IMF or Kroupa  (2001) IMF (e.g., Wilkins et al. 2008).
Our fiducial model's result (solid line) uses the VD08 IMF, and we also
show the  results with $R$ taken from  the C03 IMF.  We  find that the
VD08 IMF yields better agreement with the data, and this is partly due
to  the fact  that the  stellar remnant  mass is  lower with  this IMF
(i.e., higher $R$).

\section{Conclusions}
\label{sec:concl}

In this paper,  we use a semi-analytical model  of galaxy formation to
study  the evolution  of the  galaxy stellar  mass function,  $B$- and
$K$-band  luminosity functions, the  star formation  rate-stellar mass
relation, and the  cosmic star formation rate history.   We modify our
previous model to include cold gas accretion, an evolving stellar IMF,
and  stellar stripping  from satellites.  We focus  on the  impacts of
these effects and assumptions on  the predictions of the model, and we
obtain the following results:

\begin{itemize}

\item Cold accretion in massive  halos at high redshifts is crucial to
  produce a sufficient  number of massive galaxies at  $z>2$, in order
  to  reconcile the  model  with  the data.   By  accounting for  cold
  accretion, as well as stellar stripping from satellite galaxies, our
  model can reproduce the mild  evolution of the stellar mass function
  from  $z=1$  to  $z=0$.   This  resolves  the  problems  of  previous
  semi-analytical  models,  which   have  predicted  too  few  massive
  galaxies at high redshift and too strong evolution at low redshift.

\item The  predicted star  formation rate -  stellar mass  relation in
  previous semi-analytical  models is too  low at $z>1$. We  show that
  such  a   correlation  can  be   predicted  from  a  SAM,   and  the
  normalization can be boosted to match the data if an evolving IMF is
  adopted. The  main cause of  the higher normalization is  the larger
  number  of high-mass  stars in  our adopted  IMF (van  Dokkum 2008),
  which produces a lower stellar remnant mass, due to stronger stellar
  winds and supernovae ejecta. We find that at $z=2$ the specific star
  formation  rate is  increased by  60\% compared  to a  model  with a
  Chabrier (2003) IMF.

\item Our model is capable  of reproducing the evolution of the cosmic
  star formation rate.  An additional  constraint can be placed on the
  model by using  the number density of ULIRGs,  which have relatively
  high star  formation rates.  We find that  the combined  effect from
  cold accretion  and a bottom-light (or top-heavy)  IMF can reproduce
  the  number  density  of  ULIRGs.   The  contradiction  between  the
  measured stellar mass density and the integrated one from the cosmic
  star formation rate can be  resolved by using an IMF containing more
  high-mass stars.
\end{itemize}

In  summary, we  argue that  there are  currently no  severe conflicts
between the  CDM model  and galaxy observations.   We can  now explain
issues that were previously considered to be major problems: the rapid
formation  of  massive  galaxies  at  high  redshift  and  their  mild
evolution  at  low  redshift;  the  high  amplitude  of  the  observed
SFR-$M_{\ast}$  relation;  and the  apparent  discrepancy between  the
evolving star  formation rate density and stellar  mass density. These
discrepancies can be resolved in the context of galaxy formation based
on the CDM theory.

Finally, we note that there are some uncertainties regarding our model
assumptions.  First,  although the  tidal  stripping  of satellite  is
crucial  to stop  the rapid  growth of  massive central  galaxies, its
efficiency should  be studied in detail by  more realistic simulations
(e.g.,  Puchwein  et al.  2010).  Second,  both  the SFR-stellar  mass
relation and  evolving stellar mass  density are dependent on  the gas
recycle fraction,$R$,  which in principle should be  determined by the
stellar IMF.  Larger $R$  will decrease the  stellar remnant  mass and
increase   the  normalization  of   the  predicted   SFR-stellar  mass
relation. Due to the uncertainties  of modeling $R$ from stellar winds
and  evolution tracks,  the  favored bottom-light  (or top-heavy)  IMF
deserves further investigation.

\section{Acknowledgments}

We  are grateful  to  XianZhong Zheng,  Frank  C.  van  den Bosch  for
enlightening discussions.   We thank  the referee for  suggestions and
comments to strengthen the  presentation of our paper.  The simulation
was done  at Shanghai Supercomputer  center by the support  of Chinese
National 863 project (No.  2006AA01A125).  XK is supported by the {\it
  One Hundred Talents} project of the Chinese Academy of Science.  WPL
acknowledge  the  supports  from  Chinese National  973  project  (No.
2007CB815401),  NSFC project (No.   10873027, 10821302,  10533030) and
the Knowledge  innovation Program of  the Chinese Academy  of Sciences
(grant KJCX2-YW-T05).

%%%%%%%%%%%%%%%
% Bibliography
%%%%%%%%%%%%%%%

%%%%%%%%%%%%%%%
% Appendices
%%%%%%%%%%%%%%%

\label{lastpage}
\end{document}